# Efficient Dual-Band Single-Port Rectifier for RF Energy Harvesting at FM and GSM Bands

Nastouh Nikkhah
RF and Communication Technologies
(RFCT) research laboratory, Faculty of
Engineering and IT
University of Technology Sydney
Ultimo, NSW 2007, Australia
*Food Agility CRC Ltd.*
175 Pitt St, NSW 2000, Australia
Nastouh.Nikkhah@student.uts.edu.au

Rasool Keshavarz
RF and Communication Technologies
(RFCT) research laboratory, Faculty of
Engineering and IT
University of Technology Sydney
Ultimo, NSW 2007, Australia
Rasool.Keshavarz@uts.edu.au

Mehran Abolhasan
RF and Communication Technologies
(RFCT) research laboratory, Faculty of
Engineering and IT
University of Technology Sydney
Ultimo, NSW 2007, Australia
Mehran.Abolhasan@uts.edu.au

Justin Lipman
RF and Communication Technologies
(RFCT) research laboratory, Faculty of
Engineering and IT
University of Technology Sydney
Ultimo, NSW 2007, Australia
Justin.Lipman@uts.edu.au

Negin Shariati
RF and Communication Technologies
(RFCT) research laboratory, Faculty of
Engineering and IT
University of Technology Sydney
Ultimo, NSW 2007, Australia
*Food Agility CRC Ltd.*
175 Pitt St, NSW 2000, Australia
Negin.Shariati@uts.edu.au

*Abstract*—This paper presents an efficient dual-band rectifier for radiofrequency energy harvesting (RFEH) applications at FM and GSM bands. The single-port rectifier circuit, which comprises a 3-port network, optimized T-matching circuits and voltage doubler, is designed, simulated and fabricated to obtain a high RF-to-DC power conversion efficiency (PCE). Measurement results show PCE of 26% and 22% at −20 dBm, and also 58% and 51% at −10 dBm with a maximum amount of 69% and 65% at −2.5 dBm and −5 dBm, with single tone at 95 and 925 MHz, respectively. Besides, the fractional bandwidth of 21% at FM and 11% at GSM band is achieved. The measurement and simulation results are in good agreement. Consequently, the proposed rectifier can be a potential candidate for ambient RF energy harvesting and wireless power transfer (WPT). It should be noted that a 3-port network as a duplexer is designed to be integrated with single-port antennas which cover both FM and GSM bands as a low-cost solution. Moreover, based on simulation results, PCE has small variations when the load resistor varies from 10 to 18 kΩ. Therefore, this rectifier can be utilized for any desired resistance within the range, such as sensors and IoT devices.

I. INTRODUCTION

Radiofrequency energy harvesting (RFEH) technology has been notably considered as it enables devices to be used easily in environments where it would not be feasible to replace batteries. Moreover, RFEH can diminish environmental pollution by reducing battery production [1]. Consequently, RFEH system can be functional in self-sustainable low power applications, such as the Internet of Things (IoT) devices which are used in many areas comprising the manufacturing industry, agriculture, healthcare, wireless sensor networks (WSN), radio frequency identification devices (RFID) and intelligent transportation systems [2-5].

Radiofrequency energy harvesting from ambient signals can transfer power through free space. Since RF signals such as FM (88-108 MHz) and cellular (GSM-900) are accessible broadly and perpetually in the environment, RFEH systems are a promising solution to supply power for electronic devices. Previous research has demonstrated the feasibility of RF energy scavenging through RF field investigations in Australia [6]. FM signals can be useful for EH due to providing stable, continuously and lower pass loss in free space ambient RF source with an appropriate power level than other RF signals at different locations. In this regard, some works have been concentrated on the lower side of the RF spectrum [7-9]. In addition, GSM-900, which supports cellular communications, due to having a high number of based stations in urban and suburb areas, can be a good candidate for EH.

Rectifying antenna (rectenna) can be used to convert RF signals to DC power in RFEH scenario. A rectenna includes an antenna as an RF power receiver and a rectifier circuit to convert RF to DC power. Since ambient RF signals propagate in any direction, using antennas with omnidirectional patterns can be a suitable choice for RFEH systems [10-12]. Furthermore, the ambient RF power level received by a rectenna can alter unpredictably, depending on various factors such as power source level, distance from RF stations and RF source polarization. Consequently, the RF-to-DC conversion efficiency (PCE) is a crucial parameter in rectifier design [7]. Although some rectifiers have gained high RF to DC conversion efficiency, they are not suitable for ambient RFEH since these structures can harvest high input power levels. Comparatively, several rectennas have been designed using different kinds of rectifier topology, such as single series rectifier [13-15], Villard voltage doubler [16], differential-output topology rectifier [17], full-wave Greinacher [18] and voltage doubler rectifier [19, 20]. It should be noted that the voltage doubler topology can be well utilized in low power rectification design [2].

Since there are typical single-port antennas that resonate with both FM and GSM bands [21-23], a single-port rectifier can be an appropriate candidate to be easily integrated with these antennas. For this purpose, a 3-port network that operates as a duplexer is required to obtain a single-port rectifier. In this case, the rectifier harvests without the need for a power divider device and an extra

SMA connector. Consequently, the proposed system is low-cost, compact and can be fabricated easily.

This paper describes an efficient dual-band rectifier for RF energy harvesting operating over a single port in FM and GSM bands. The proposed RF energy harvester comprises a 3-ports network, optimal T-matching circuits including radial stubs and a voltage doubler structure which is designed, simulated and measured. The results of the rectifier demonstrate a noticeable fractional bandwidth where the reflection coefficient is defined below –10 dBm. Furthermore, the high conversion efficiency is achieved over a proposed input power range. Further, the efficiency is relatively stable for variations of the load resistor. Hence, this system can be used for any desired sensor with resistance between 10 to 18 kΩ. The rest of the paper is organized as follows: Section II describes the proposed rectifier design and presents simulation and measurement results. Finally, Section III concludes this paper.

## II. RECTIFIER DESIGN, SIMULATION AND MEASUREMENT

The block diagram of the proposed RF energy harvester is illustrated in Fig. 1, which comprises an antenna as an RF signal receiver, a 3-port network to obtain a single-port structure, a T-matching circuit to transfer maximum power from the antenna to load, a voltage doubler to convert input RF energy (AC signal) to DC voltage and a load section. The proposed single-port dual-band rectifier circuit, which harvests energy from FM and GSM bands simultaneously, is presented in Fig 2. It should be noted that the rectifier design is simulated and optimized using Advance Design System (ADS) software.

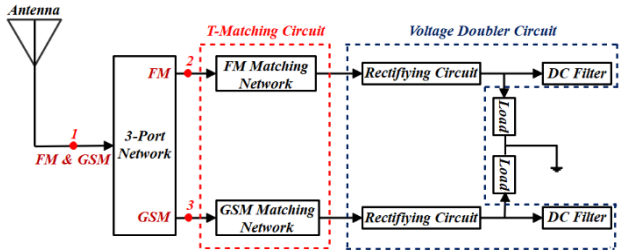

Fig.1. Block diagram of the proposed RF energy harvester.

Low turn-on voltage Schottky diode HSMS-2850 ($C_{j0}$ = 0.18 pF, $R_S$ = 25 Ω, $I_S$ = 3e−6 A) is chosen as a rectification device in voltage doubler structure, which is used especially in a small signal application at frequency bellow 1.5 GHz due to its low series resistance ($R_S$) and also junction capacitance ($C_{j0}$) [24]. In addition, the HSMS-2850 diode has a low threshold voltage with high saturation current, which is a critical factor for rectification at low input power levels [2].

Different rectifiers topologies have been proposed, such as bright type, single series, single shunt, and voltage doubler rectifiers. A voltage doubler rectifier is used due to its desired advantages, including, first, the total amplitude of the output voltage is boosted since two diodes are added in series. Second, using a voltage doubler rectifier decreases diodes' junction resistance and enhances the detection sensitivity at low power levels as the produced current of the first diode provides external bias for the second diode [25].

A resistance load is added at the end of the voltage doubler to sense the DC voltage. Since the resistance values of applied sensors are relatively near to the range of 10 to 18 kΩ [20], this range is chosen for optimization in ADS software at low input power levels to maximize the RF to DC power conversion efficiency. In this case, 14 kΩ was achieved as an optimum value for the load resistance.

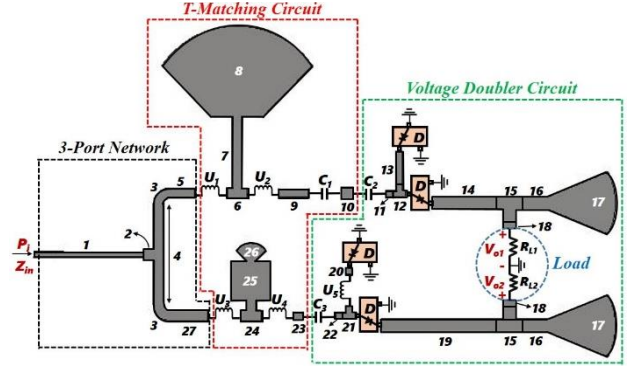

Fig.2. Topology of the proposed rectifier. Dimensions (mm), inductors (nH), capacitors (pF) and loads (kΩ) are presented in Table I.

TABLE I. PARAMETERS OF THE PROPOSED STRUCTURE SHOWN IN FIG. 2.

| [$W_1$] | [$W_{2,3}$] | [$R_3$] | [$W_4$] | [$W_5$] | [$W_6$] | [$W_7$] | [$W_8$] |
|---|---|---|---|---|---|---|---|
| 0.5 | 1.6 | 1.6 | 1.6 | 1.6 | 1.5 | 1.5 | 8 |
| [$W_9$] | [$W_{10}$] | [$W_{11}$] | [$W_{12}$] | [$W_{13}$] | [$W_{14}$] | [$W_{15}$] | [$W_{16}$] |
| 1.1 | 2 | 1 | 1 | 1 | 2 | 2 | 2 |
| [$W_{17}$] | [$W_{18}$] | [$W_{19}$] | [$W_{20}$] | [$W_{21}$] | [$W_{22}$] | [$W_{23}$] | [$W_{24}$] |
| 2 | 2 | 2 | 1 | 1 | 1.1 | 1.2 | 1.6 |
| [$W_{25}$] | [$W_{26}$] | [$W_{27}$] | [$L_1$] | [$L_4$] | [$L_5$] | [$L_7$] | [$L_8$] |
| 6 | 1 | 1.6 | 18 | 15.8 | 7.6 | 9.6 | 11 |
| [$L_9$] | [$L_{10}$] | [$L_{11}$] | [$L_{13}$] | [$L_{14}$] | [$L_{16}$] | [$L_{17}$] | [$L_{18}$] |
| 4.5 | 2 | 1 | 5 | 5 | 4 | 10 | 1 |
| [$L_{19}$] | [$L_{20}$] | [$L_{22}$] | [$L_{23}$] | [$L_{25,27}$] | [$L_{26}$] | $Ang_{8,26}$ | $Ang_{17}$ |
| 32.4 | 1 | 1.1 | 1.5 | 6 | 3.4 | 90° | 45° |
| $U_1$ | $U_2$ | $U_3$ | $U_4$ | $U_5$ | $C_{1,2}$ | $C_3$ | $R_L$ |
| 230 | 1970 | 3.2 | 55 | 3 | 20 | 100 | 14 |

Furthermore, the impedance matching network section is optimized based on T-circuits, allowing the upper stage of the rectifier to receive the maximum FM signal from an input port. At the same time, the lower one can capture GSM band signals. It is noteworthy to mention that a radial stub used in the matching section with different dimensions can maintain the rectifier performance over different load resistors. Moreover, the radial stubs (parts 8, 26 and 17 shown in Fig. 2) were used instead of capacitors, simplifying the rectifier design as any desired values can be achieved at a low cost. In addition, using the radial stub at the end of the rectifier as a capacitor role (part 17) can save energy due to the imaginary input impedance of the transmission line and smooth the output DC signal as a low pass filter.

A 3-port network that operates similar to a duplexer is designed to efficiently connect the output port of any desired single-port RF receiver ($Z_{in}$=50 Ω), which supports both FM and GSM bands, to both rectifier stages. As shown in Fig.1 and Table II, much of the input signal transmits through the upper stage since it has a notably higher transmission coefficient ($S_{21}$) than the lower stage at the FM band. On the other hand, the lower stage transfers GSM signals due to the significant transmission coefficient ($S_{31}$). The scattering parameters in both bands were simulated over different input powers.

TABLE II. S-PARAMETERS OF 3-PORT NETWORK SHOWN IN FIG.1.

| Frequency band | Input power (dBm) | $S_{11}$ (dB) | $S_{21}$ (dB) | $S_{31}$ (dB) | $S_{23}$ (dB) |
|---|---|---|---|---|---|
| FM | −2.5 | −24 | −0.14 | −16 | −16.6 |
|  | −5 | −24.1 | −0.13 | −16.2 | −16.8 |
|  | −10 | −24 | −0.11 | −17.1 | −17.6 |
|  | −15 | −23.5 | −0.10 | −18 | −18.4 |
| GSM | −2.5 | −10.2 | −40 | −0.47 | −37.7 |
|  | −5 | −17.1 | −37.7 | −0.12 | −37.7 |
|  | −10 | −17 | −37.1 | −0.13 | −37.7 |
|  | −15 | −14.7 | −35.9 | −0.19 | −37.7 |

The proposed rectifier is fabricated on RO4003C with a dielectric constant of $\varepsilon_r = 3.38$, loss tangent of $\tan\delta = 0.0027$ and thickness of $t_s = 0.8$ mm as shown in Fig.3. The optimized parameters of the rectifier are listed in Table I.

Fig. 4 (a) and Fig. 4 (b) depict the rectifier's simulated and measured reflection coefficient at different input power levels from −40 to +10 dBm with the step of about 2.5 dBm. Reasonable agreement between simulation and measurement results is achieved at both frequency bands. Moreover, the simulation and measurement reflection coefficients at FM and GSM frequency bands are illustrated in Fig. 4 (c) and Fig. 4 (d). In order to demonstrate the impedance matching bandwidth of the rectifier at desired frequency bands, the corresponding power levels to the minimum reflection coefficient are chosen as input power (−2.5 dBm at 95 MHz and −5 dBm at 925 MHz based on Fig.4). Also, Fig.4 (c) and Fig. 4 (d) show the fractional bandwidth of 21% and 11% at FM and GSM bands, respectively. Consequently, it can be observed that the proposed rectifier has an acceptable impedance matching across both frequency bands, which makes it a good candidate to be integrated with an antenna resonating at FM and GSM bands concurrently to realize a rectenna for RF energy harvesting.

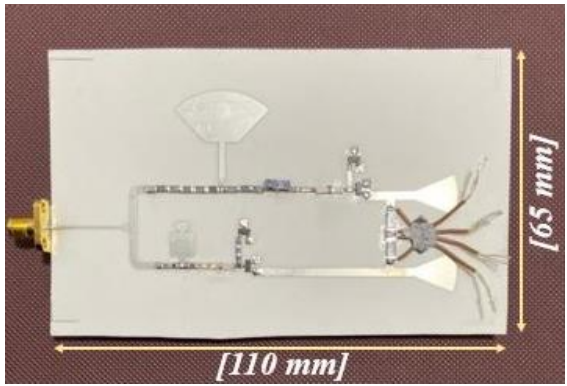

Fig.3. Fabricated rectifier prototype.

Efficiency is an important metric in energy harvesters which can be defined as RF-to-DC power conversion efficiency (PCE) and can be calculated as below,

$$\eta_{RF-DC} = \frac{V_{Odc}^2}{P_i\, R_L} \times 100\% \qquad (1)$$

Where $V_{Odc}$ is the output DC voltage in each stage, $R_L$ is the load resistor, and $P_i$ is the input power provided by the signal generator to the rectifier. The evaluation of RF-to-DC efficiency and output DC voltage are demonstrated in Fig 5. The measured results approximately agree with the simulations in both bands. It should be noted that ohmic loss, specifically due to using radial stubs at T- matching part and output of the circuit instead of a passive element, can lead to a small difference between simulation and measurement results. Based on the measurement results, the proposed rectifier demonstrates an RF-to-DC efficiency of about 26% and 21% at $P_{in}= -20$ dBm, 58% and 51% at $P_{in}= -10$ dBm, with a maximum PCE of 69% and 65%, at $P_{in}= -2.5$ and $-5$ dBm at FM and GSM bands, respectively. For further clarification, the measurement outputs are also presented in Table III. Furthermore, the simulation results at fixed frequencies of 95 and 925 MHz in Fig. 6 indicate the relation between RF-to-DC efficiency and $R_L$ over different input power levels. $R_L$=14 kΩ is an optimum load for the proposed rectifier, leading to provide higher efficiencies than other loads in both frequency bands. Also, Fig. 6 shows that the proposed rectifier performance achieves a relatively light variation of PCE over the range of 10 to 18 kΩ. Consequently, the proposed rectifier can be a proper candidate for ambient RF energy harvesting since it presents a dual-band function with an adequate efficiency at input power levels such as −20 dBm (10 µW).

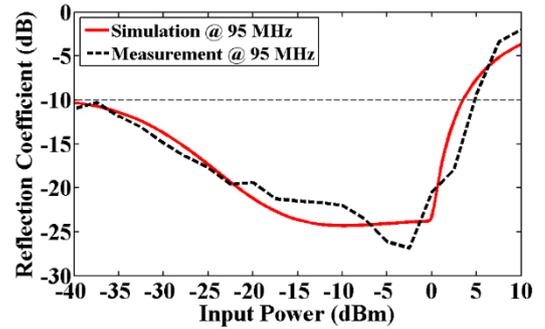

(a)

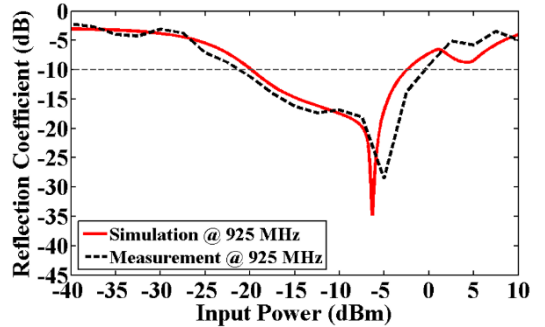

(b)

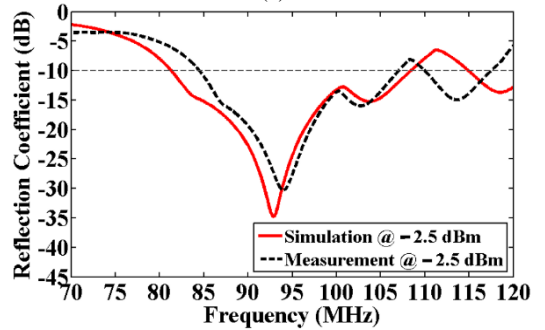

(c)

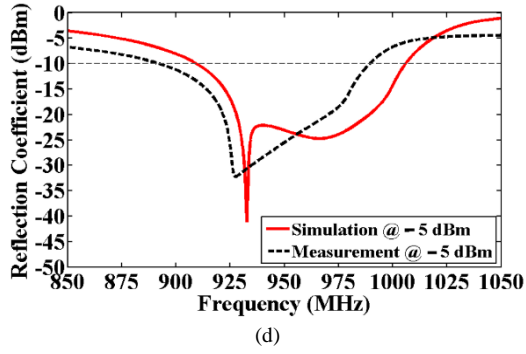

(d)

Fig.4. Simulated and measured reflection coefficient of the proposed rectifier at different input RF power levels from −40 to +10 dBm (a) FM (95MHz), (b) GSM (925MHz) frequency band and (c) FM band at −2.5dBm, (d) GSM band at −5dBm.

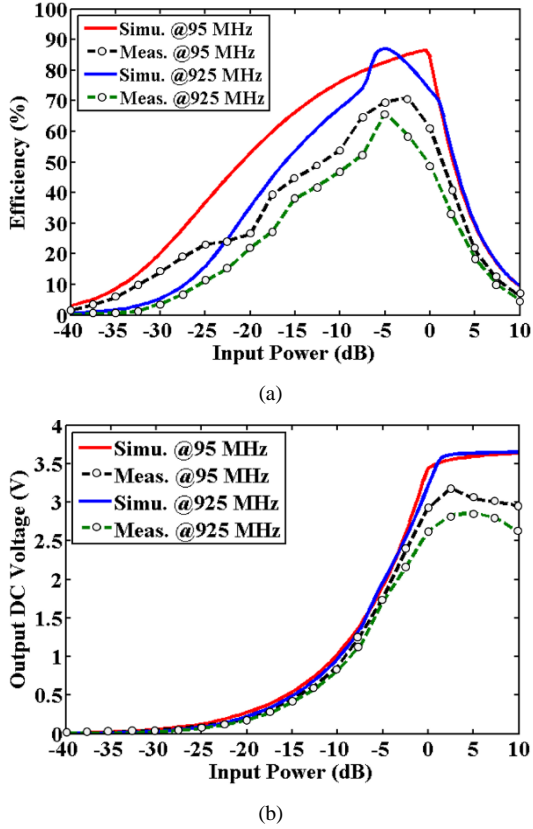

(a)

(b)

Fig.5. Simulated and measured results of the proposed rectifier over different input RF power levels at FM and GSM frequency bands with 14 kΩ load, (a) RF-to-DC conversion efficiency and (b) output DC voltage.

TABLE III. THE MOST IMPORTANT MEASUREMENT RESULTS.

| Frequency band | Input power (dBm) | RF-to-DC conversion efficiency at a single tone | $V_{Odc}$ (V) |
|---|---|---|---|
| FM | −2.5 | 69% | 2.33 |
| | −10 | 58% | 0.901 |
| | −20 | 26% | 0.191 |
| GSM | −5 | 65% | 1.69 |
| | −10 | 51% | 0.845 |
| | −20 | 21% | 0.171 |

Finally, Table IV compares recently reported rectifiers with the proposed system. It is evident from Table IV that this work has achieved higher fractional bandwidth and RF-to-DC efficiency at a single tone compared to other works with equal input powers. The obtained fractional bandwidths are 21% and 11% at FM and GSM bands, respectively. In addition, as mentioned before, high efficiency is the critical feature of RF energy harvesters. The proposed rectifier achieves an RF-to-DC conversion efficiency of 58% and 51% at -10 dBm and also 26% and 22% at a low input power of -20 dBm at FM and GSM bands, respectively. Hence, the proposed rectifier can be a suitable candidate for ambient RF energy harvesting and WPT systems.

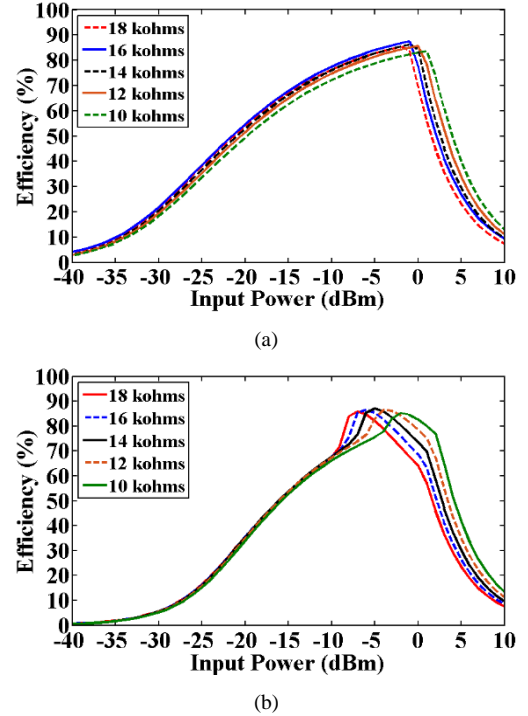

(a)

(b)

Fig.6. Simulated RF-to-DC conversion efficiency of the proposed rectifier for five samples of load ($R_L$) over different input RF power levels at (a) FM (95 MHz) and (b) GSM (925 MHz) frequency bands.

TABLE IV. COMPARISON OF THE PROPOSED RECTIFIER AND OTHER REPORTED WORKS.

| Ref. | Frequency (MHz) | Fractional bandwidth (%) | RF-to-DC conversion efficiency at a single tone |
|---|---|---|---|
| [2] | 490 (1), 860 (2) | 4 (1), 2 (2) | 17% @ −10 dBm (1) <br> 7% @ −10 dBm (2) |
| [9] | 88 - 108 | 23 | 40% @ −10 dBm <br> 18% @ −20 dBm |
| [14] | 925 (1), 1820 (2), 2170 (3) | 4 (1), 5 (2), 3 (3) | 42% @ −10 dBm at (1) <br> 25% @ −20 dBm at (1) <br> 32% @ −10 dBm at (2) <br> 18% @ −20 dBm at (2) <br> 25% @ −10 dBm at (3) <br> 13% @ −20 dBm at (3) |
| [15] | 1840 | 7 | 37% @ −10 dBm <br> 21.1% @ −20 dBm |
| [21] | 97.5 (1), 868 (2) | 11 (1), 8 (2) | 46% @ −10 dBm at (1) <br> 27% @ −20 dBm at (1) <br> 29% @ −10 dBm at (2) <br> 14% @ −20 dBm at (2) |
| [26] | 2450 (1), 5800 (2) | 4 (1), 2 (2) | 20% @ −10 dBm at (1) <br> 3% @ −20 dBm at (1) <br> 22% @ −10 dBm at (2) <br> 3% @ −20 dBm at (2) |
| This work | 95 (1), 925 (2) | 21 (1), 11 (2) | 58% @ −10 dBm at (1) <br> 26% @ −20 dBm at (1) <br> 51% @ −10 dBm at (2) <br> 22% @ −20 dBm at (2) |

## III. Conclusion

This paper demonstrates an efficient dual-band rectifier to harvest RF energy from FM and GSM bands simultaneously. The single-port dual-band voltage doubler rectifier was designed, simulated and fabricated, which presents a good impedance matching performance with fractional bandwidth of 21 and 11% at FM and GSM bands, respectively, using optimized 3-port network and T-matching circuits. Also, the proposed energy harvester is a low-cost solution for single-port antennas that resonate at both FM and GSM bands due to not using the power divider device. Measurement results of the rectifier show a satisfactory efficiency over different low input power levels. The proposed rectifier obtains the RF-to-DC efficiency of 58% at 95 MHz and 51% at 925 MHz when the power input level is −10 dBm, respectively. The achieved efficiencies at FM and GSM bands and over a broad input power ranges make the proposed rectifier a good candidate for ambient RF energy harvesting and wireless power transmission systems.

## Acknowledgment

This project was supported by funding from Food Agility CRC Ltd, funded under the Commonwealth Government CRC Program. The CRC Program supports industry-led collaborations between industry, researchers and the community.

## References

[1] S. Kim, R. Vyas, J. Bito, K. Niotaki, A. Collado, A. Georgiadis, and M. M. Tentzeris, "Ambient RF Energy-Harvesting Technologies for Self-Sustainable Standalone Wireless Sensor Platforms," Proc. IEEE, vol. 102, no. 11, pp. 1649–1666, November 2014.

[2] N. Shariati, W. S. T. Rowe, J. R. Scott, and K. Ghorbani, "Multi-service highly sensitive rectifier for enhanced RF energy scavenging," Scientific Reports, vol. 7, May 2015.

[3] I. Zhou, I. Makhdoom, N. Shariati, M. A. Raza, R. Keshavarz, J. Lipman, M. Abolhasan and A. Jamalipour, "Internet of Things 2.0: Concepts, Applications, and Future Directions," IEEE Access, vol. 9, pp.70961–71012, May 2021.

[4] M. A. Ullah, R. Keshavarz, M. Abolhasan, J. Lipman, K. P. Esselle and N. Shariati, "A Review on Antenna Technologies for Ambient RF Energy Harvesting and Wireless Power Transfer: Designs, Challenges and Applications," IEEE Access, vol. 10, pp. 17231-17267, February 2022.

[5] A. Raza, R. Keshavarz and N. Shariati, "Miniaturized Patch Rectenna Using 3-Turn Complementary Spiral Resonator for Wireless Power Transfer," IEEE Asia-Pacific Microwave Conference (APMC), Australia, pp. 455-457, November 2021.

[6] N. Shariati, W.S.T. Rowe, and K. Ghorbani, "RF Field Investigation and Maximum Available Power Analysis for Enhanced RF Energy Scavenging," 42nd EuMC. Netherlands, pp.329–332, October 2012.

[7] N. Shariati, W.S.T. Rowe, and K. Ghorbani, "Highly Sensitive FM Frequency Scavenger Integrated in Building Materials," 45th EuMC. France, pp.68–71, September 2015.

[8] A. Noguchi, and H. Arai, "Small loop rectenna for RF energy harvesting," APMC., South Korea, November 2013.

[9] E. M. Jung, Y. Cui, T. H. Lin, X. Hi, A. Eid, J. G. D. Hester, G. D. Abowd, T. E. Starner, W. S. Lee, and M. M. Tentzeris, "A Wideband, Quasi-Isotropic, Kilometer-Range FM Energy Harvester for Perpetual IoT," IEEE Microwave and Wireless Components Letters, vol.30, pp. 201 – 204, February 2020.

[10] N. Nikkhah, and B. Zakeri, "Efficient design and implement an electrically small HF antenna," IEEE 4th KBEI. Iran, December 2017.

[11] N. Nikkhah, B. Zakeri, and H. Abedi, "Extremely electrically small MF/HF antenna," IET Microwaves, Antennas & Propagation, vol. 14, pp.88–92, January 2020.

[12] M. Rad, N. Nikkhah, B. Zakeri, and M. Yazdi, "Wideband Dielectric Resonator Antenna with Dual Circular Polarization," IEEE Transaction on Antennas and Propagation, vol. 70, pp. 714–719, January 2022.

[13] H. Sun, Y. X. Guo, M. He, and Z. Zhong, Noble, "A Dual-Band Rectenna Using Broadband Yagi Antenna Array for Ambient RF Power Harvesting," IEEE Antennas Wireless Propagation Letter, vol.12, pp. 918 – 921, July 2013.

[14] S. Shen, C. Y. Chiu, and R. D. Murch, Noble, "A Dual-Port Triple-Band L-Probe Microstrip Patch Rectenna for Ambient RF Energy Harvesting," IEEE Antennas Wireless Propagation Letter, vol.16, pp. 3071 – 3074, October 2017.

[15] S. Shen, C. Y. Chiu, and R. D. Murch, Noble, "Multiport Pixel Rectenna for Ambient RF Energy Harvesting," IEEE Transactions on Antennas and Propagation, vol.16, pp. 644 – 656, February 2018.

[16] S. Chandravanshi, S. S. Sarma, and M. J. Akhtar, "Design of Triple Band Differential Rectenna for RF Energy Harvesting," IEEE Transaction on Antennas and Propagation, vol.66, pp. 2716 – 2726, March 2018.

[17] V. Kuhn, C. Lahuec, F. Seguin, and C. Person, "A Multiband Stacked RF Energy Harvester With RF-to-DC Efficiency Up to 84%," IEEE Transactions On Microwave Theory and Techniques, vol.63, pp. 1768 – 1778, May 2015.

[18] M. Zeng, A. S. Andrenko, X. Liu, Z. Li, and H. Z. Tan, "A Compact Fractal Loop Rectenna for RF Energy Harvesting," IEEE Antennas Wireless Propagation. Letter, vol.16, pp. 2424 – 2427, July 2017.

[19] Y. Shi, Y. Fan, Y. Li, L. Yang, and M. Wang, "An Efficient Broadband Slotted Rectenna for Wireless Power Transfer at LTE Band," IEEE Transaction on Antennas and Propagation, vol.67, pp. 814 – 822, February 2019.

[20] R. Keshavarz, and N. Shariati, "Highly Sensitive and Compact Quad-Band Ambient RF Energy Harvester," IEEE Transactions on Industrial Electronics, vol. 69, pp. 3609-3621, April 2021.

[21] S. N. Daskalakis, A. Georgiadis, A. Bletsas, and C. Kalialakis, "Dual band RF harvesting with low-cost lossy substrate for low-power supply system," 42th EuCAP, Switzerland, April 2016.

[22] I. A. Osaretin, A. Torres, and C. C. Chen, "A Novel Compact Dual-Linear Polarized UWB Antenna for VHF/UHF Applications," IEEE Antennas Wireless Propagation. Letter, vol.8, pp. 145 – 148, January 2009.

[23] Y. Liu, Z. Ai, G. Liu, and Y. Jia, "An Integrated Shark-Fin Antenna for MIMO-LTE, FM, and GPS Applications," IEEE Antennas Wireless Propagation. Letter, vol.18, pp. 1666 – 1670, August 2009.

[24] Avago Technologies. Surface Mount Zero Bias Schottky Detector Diodes HSMS-285x Series. Available: http://www.avagotech.com.

[25] N. Shariati, W.S.T. Rowe, and K. Ghorbani, "Highly sensitive rectifier for efficient RF energy harvesting," 44th EuMC. Italy, October 2014.

[26] M. Rehman, W. Ahmad, and W. T. Khan, "Highly efficient dual band 2.45/5.85 GHz rectifier for RF energy harvesting applications in ISM band," Proc. IEEE Asia Pacific Microw. Conf., Malaysia, pp. 150–153, January 2018.